\documentclass[aa]{aa}

\usepackage[rightcaption]{sidecap}
\usepackage[varg]{txfonts}
\usepackage{graphicx}
\usepackage{natbib}
\usepackage{booktabs}

\usepackage{multirow}
\usepackage{amsmath}
\usepackage{amssymb}
\usepackage{adjustbox}
\usepackage{url}
\usepackage{appendix}
\usepackage{xcolor}
\usepackage{hyperref}
\bibpunct{(}{)}{;}{a}{}{,} 
\usepackage{cleveref}
\let\VANthebibliography\thebibliography
\def\thebibliography{\DeclareRobustCommand{\VAN}[3]{##3}\VANthebibliography}

\newcommand{\cenx}{Cen\,X-3}
\newcommand{\vstar}{V779\,Cen}
\newcommand{\xrism}{XRISM/Resolve}
\newcommand{\feka}{Fe~K$\alpha$}

\begin{document}

\title{Spectral Analysis of the Egress and Ingress Phases of \cenx\ with \xrism}

\titlerunning{XRISM/Resolve spectroscopy of \cenx}

\author{J. Planelles-Villalva$^{1}$, 
L. Gu$^{2}$, J.J. Rodes-Roca$^{1}$, G. Sanjurjo-Ferr\'{i}n$^{1}$, 
J. de Plaa$^{2}$, J.M. Torrej\'on$^{1}$, 
F. S. Porter$^{3}$, 
Y. Mochizuki$^{4,5}$, 
M. Loewenstein$^{3,6,7}$,  
E. Costantini$^{2,8}$} 
\authorrunning{Planelles-Villalva} %

\institute{
$^{1}$Instituto Universitario de F\'{i}sica Aplicada a las Ciencias y las Tecnolog\'ias, Universidad de Alicante, 03690 Alicante, Spain\\
\email{jessica.planelles@ua.es}\\
$^{2}$SRON Space Research Organisation Netherlands, Niels Bohrweg 4, 2333 CA Leiden, The Netherlands\\
$^{3}$NASA/Goddard Space Flight Center, 8800 Greenbelt Rd, Greenbelt, MD 20771 USA \\
$^{4}$Department of Astronomy, Graduate School of Science, The University of Tokyo, 7-3-1 Hongo, Bunkyo-ku, Tokyo 113-0033, Japan\\
$^{5}$Institute of Space and Astronautical Science (ISAS), Japan Aerospace Exploration Agency (JAXA), 3-1-1 Yoshinodai, Chuo-ku, Sagamihara, Kanagawa 252-5210,
Japan\\
$^{6}$Department of Astronomy, University of Maryland, College Park,
MD 20742, USA\\
$^{7}$ Center for Research and Exploration in Space Science and Technology, NASA / GSFC (CRESST II), Greenbelt, MD 20771, USA\\
$^{8}$Anton Pannekoek Institute for Astronomy, University of Amsterdam, Science Park 904, NL-1098 XH Amsterdam, The Netherlands
}

\date{}

\abstract
{}
{We investigate the phase-resolved X-ray spectral evolution of the high-mass X-ray binary Cen\,X-3 across the eclipse ingress and egress transitions, using two out-of-eclipse intervals as references, to constrain the properties of the absorbing material and the Fe\,K emitting regions with high-resolution XRISM/Resolve spectroscopy.}
{The spectra are modeled with SPEX using a multi-component model consisting of a fixed interstellar absorber, a local wind absorber with free temperature and velocity broadening, a power-law continuum, a blackbody component, and a disk-reflection component (\texttt{xskirtor}), which accounts for the fluorescent Fe\,K$\alpha$ and K$\beta$ emission lines. Additional Gaussian components are included to model the ionized Fe\,\textsc{xxv} and Fe\,\textsc{xxvi} emission lines. We compare the orbital variation of the absorbing column with the predictions of a smooth CAK wind model and investigate the orbital variability of the Fe\,K$\alpha$ emission and its velocity broadening.}
{The local absorbing column density varies by nearly two orders of magnitude over the orbit, from $N_\mathrm{H}^\mathrm{local}=4.25\times10^{22}$\,cm$^{-2}$ during pre-ingress to $5.9\times10^{23}$\,cm$^{-2}$ at ingress. The excess absorption observed during ingress cannot be reproduced by a smooth stellar-wind model alone, suggesting the presence of a localized overdense structure along the line of sight. The Fe\,K$\alpha$ line width also varies with orbital phase, from $\sim510$\,km\,s$^{-1}$ during pre-ingress to $\sim1400$\,km\,s$^{-1}$ during egress, with intermediate, statistically indistinguishable values of $\sim900$\,km\,s$^{-1}$ measured during ingress and post-egress. The highly ionized Fe\,\textsc{xxv} and Fe\,\textsc{xxvi} emission is detected throughout the orbit, although fewer individual line components are resolved during ingress. Together, the absorption, fluorescence, and ionized emission diagnostics reveal a complex circumstellar environment in Cen\,X-3, with localized dense material and multiple reprocessing regions contributing to the observed orbital variability.}
{These results provide a detailed view of the absorbing, emitting, and reprocessing environment during the eclipse transitions and demonstrate the capability of XRISM/Resolve to probe the geometry and kinematics of accretion environments in eclipsing high-mass X-ray binaries.}


\maketitle

\section{Introduction}
\label{sec:introduction}

High-mass X-ray binaries (HMXBs) are among the most powerful laboratories for studying accretion physics under extreme conditions. In these systems, a compact object---typically a neutron star (NS) or black hole (BH)---accretes matter from a massive stellar companion, producing X-ray emission shaped not only by the intrinsic emission mechanism but also by complex interactions with the surrounding stellar wind, absorption, scattering, and the binary geometry \citep{1972A&A....21....1P, 2006csxs.book..623T}.

Centaurus X-3 (Cen~X-3) is one of the most extensively studied HMXBs since its discovery by the {Uhuru} satellite in 1971 \citep{1971ApJ...167L..67G}, with \cite{1972ApJ...172L..79S} subsequently identifying the 4.8\,s pulsation period. The system consists of a NS orbiting an O6--8\,III giant companion (V779\,Cen) with an orbital period of $\sim 2.087$\,days \citep{2015A&A...577A.130F}, and its nearly edge-on geometry, with an inclination of $i \simeq 73^\circ$--$79^\circ$ \citep{2024A&A...690A.360S}, produces regular X-ray eclipses well suited to orbital-phase-resolved spectroscopy of the stellar wind.

Beyond its orbital eclipses, Cen~X-3 also exhibits complex long-term variability: periodic high and low X-ray states on timescales of 125--165\,days \citep{Priedhorsky1983}, a super-orbital period of $220\pm5$\,days attributed to disk precession \citep{2022RMxAA..58..355T}, and possible contributions from circumstellar structures such as a bow shock ahead of the accretion stream \citep{Suchy2008}, making it one of the most complex binaries known. \cite{2010RAA....10.1127D} found sharper, shorter eclipse transitions in high-soft states versus shallower, longer ones in low-hard states, attributing this to varying obscuration by the precessing disk, while \cite{Liu2024} linked torque reversals to the orbital light-curve profile---together suggesting that the geometry of the accretion flow drives much of the observed variability and motivating phase-resolved spectroscopy capable of disentangling its circumstellar components.

Previous studies using \textit{Chandra}/HETGS \citep{2003ApJ...582..959W, 2005ApJ...634L.161I}, \textit{XMM-Newton}/RGS and EPIC \citep{2021MNRAS.501.5892S,2024A&A...690A.360S, Aftab2019, 2026A&A...708A.113R}, and \textit{Suzaku} \citep{Naik2011} have revealed a rich orbital-phase-dependent spectral phenomenology. During eclipse, when the direct pulsar emission is blocked, the observed spectrum is dominated by radiation scattered and reprocessed in the extended stellar wind \citep{Aftab2019}, making eclipse spectroscopy a sensitive probe of wind structure free from continuum contamination. The ingress and egress phases are of particular interest, as the line of sight progressively traverses different wind layers, revealing gradients in column density, ionization, and velocity; however, the limited spectral resolution and low sensitivity at energies above 2 keV of previous missions has often prevented detailed studies of both emission and absorption signatures during these highly absorbed phases.

The iron K$\alpha$ complex in the 6.4--7.0\,keV band provides fundamental diagnostics of the absorbing column, ionization state, and dynamics of material along the line of sight to the NS \citep{1999ApJ...525..921S, 2002ApJ...564L..21S, Kallman2004}. The fluorescent 6.4\,keV line, arising from cold or weakly ionized iron in optically thick material illuminated by the NS continuum \citep{George1991, 1991A&A...247...25M}, traces the cold, dense regions of the wind, while the associated Fe\,\textsc{i} K absorption edge at $E \simeq 7.1$\,keV, produced by photoelectric absorption in the same cold, weakly ionized material, and its shape and broadening may provide information on the kinematics of the absorbing material. The weaker, highly ionized Fe\,\textsc{xxv} and Fe\,\textsc{xxvi} lines, arising from hot photoionized plasma closer to the compact object \citep{Porquet2001}, provide a complementary diagnostic of the ionization structure, together making the Fe~K band a powerful probe of both the absorbing wind and the reprocessing regions in eclipsing HMXBs \citep{Yaqoob2023}.

The launch of the \textit{X-Ray Imaging and Spectroscopy Mission} (XRISM) in September 2023 \citep{Tashiro2025} has expanded the capabilities of high-resolution X-ray spectroscopy. The Resolve instrument provides an energy resolution of $\sim 5$\,eV (FWHM) over the 1.7--12\,keV band \citep{Ishisaki2025}, providing high-resolution spectroscopy at Fe-K energies with higher sensitivity than previous grating observations. In Cen~X-3, the He-like Fe\,\textsc{xxv} triplet was already resolved with \textit{Chandra}/HETGS by \citet{2005ApJ...634L.161I}, using a single 45\,ks exposure restricted to the post-egress phase interval ($\phi_{\rm orb} = 0.13$--$0.40$), without covering the ingress or egress transitions themselves. Previous \textit{XMM-Newton}/RGS studies of Cen~X-3 have also focused on phase-averaged eclipse and out-of-eclipse spectra \citep{2026A&A...708A.113R}, rather than resolving the rapid spectral changes across eclipse transitions. The combination of high spectral resolution and sensitivity of XRISM/Resolve therefore enables time-resolved high-resolution spectroscopy across the full eclipse ingress and egress transitions.

In this work, we use the XRISM/Resolve observation of Cen~X-3 to investigate
the physical and geometrical properties of the stellar wind across the
eclipse ingress, egress, and out-of-eclipse (OOE) orbital phases. By
performing high-resolution spectroscopy of the local absorption and the
Fe~K$\alpha$ emission line, we aim to characterize the orbital evolution of
the absorbing column density---with particular attention to the
ingress/egress asymmetry---and to investigate the velocity broadening of
the Fe~K$\alpha$ emission line, together with a test for
possible velocity broadening of the Fe~K absorption edge, as potential
probes of the kinematics of the circumstellar material, thereby probing
the interaction between the stellar wind and the accretion environment.

The paper is organized as follows. In Sect.~\ref{sec:data_reduction}, we describe the XRISM observations, the data reduction procedures, and the selected orbital-phase intervals. In Sect.~\ref{sec:results}, we present the results of the spectral analysis for each orbital phase selected. In Sect.~\ref{sec:discussion}, we discuss the physical implications of our findings in the context of the stellar wind structure of Cen~X-3. Finally, in Sect.~\ref{sec:conclusions}, we summarize our main conclusions.

\section{Observation and data}
\label{sec:data_reduction}

We analyzed the XRISM observation (sequence ID 300003010), carried out between 2024 February 12 at 23:56:04 and February 15 at 06:19:04, with a total exposure time of 196 ks. The data were obtained as part of the performance-and-verification (PV) phase.

During an initial inspection of the spectral products, we compared spectra extracted from the inner pixel set (pixels 0, 17, 18, and 35) with those from the outer pixel set (all remaining active pixels, excluding calibration pixel~12 and flagged pixel~27). In the absence of instrumental effects, both pixel groups are expected to measure the same source spectrum, and their spectral ratio should therefore remain approximately flat over the entire energy range.

\begin{figure}[h!]
    \centering
    \includegraphics[width=0.5\textwidth]{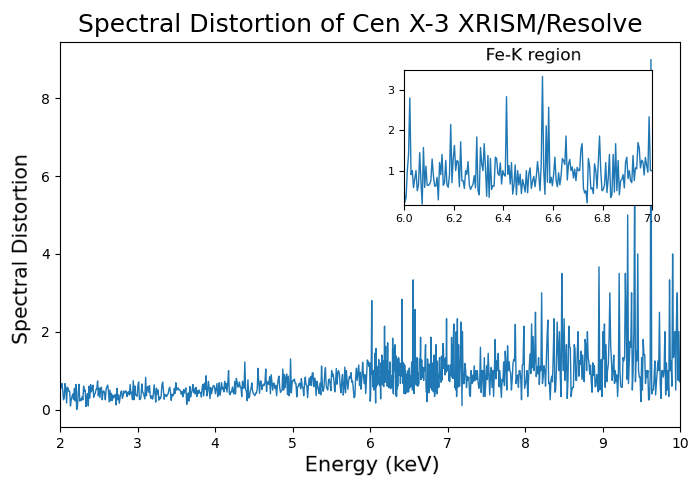}
    {Ratio of the spectra extracted from the inner pixel array
    (pixels 0, 17, 18, and 35) to those extracted from the outer pixel
    array for the \xrism\ observation of \cenx. \textit{Inset:} Zoom into
    the 6.0--7.0\,keV band.}
    \label{fig:distortion}
\end{figure}

Instead, the outer-to-inner spectral ratio exhibits an excess around the Fe~K$\alpha$ line at 6.4\,keV (Fig.~\ref{fig:distortion}), with residuals near 6.7\,keV. These features are consistent with a gain calibration error affecting primarily the inner pixel array, producing an artificial distortion of the iron K band.

\subsection{Gain calibration}
The gain of the XRISM/Resolve instrument varies with time owing to changes in the thermal environment of the detector system. To correct the time-dependent gain, and thus the time-dependent energy scale, a series of gain-monitoring measurements using on-board calibration sources are interwoven with each celestial observation \citep{2025JATIS..11d2016P}. Under normal circumstances, this procedure recovers the energy scale with high precision, yielding a systematic uncertainty generally better than 0.2\,eV at 5.9\,keV \citep{2025JATIS..11d2018E}.

For every Guest-Observer observation, the data and its calibration are carefully checked by one or more instrument scientists, and failures in the gain fitting are caught before the data is released to investigators. For the PV phase observation of Cen X-3 reported here, the same verification process was carried out by XRISM science team members as part of internal review; however, the specific gain-fitting failure affecting the inner four pixels was not identified at that stage. Now that PV phase observations are publicly released, post-facto review of a few sources, such as this one, is needed to provide a robust calibration.

The PV observation of Cen X-3 is one such case, in which the celestial source itself interferes with the recovery of the time-dependent energy scale. It is an example of a case where some fraction of the gain-monitoring measurements occurs on top of a bright source during times when the source is not occulted by the Earth. As a result, many calibration-source events migrate from high-resolution to lower grades, decreasing the statistics in the high-resolution spectra used for gain monitoring. This only happens for a handful of point sources and generally only affects, for a source centered in the array, the central four pixels where the largest fraction of the events are detected. When this happens, the gain recovery portion of the XRISM/Resolve pipeline must be carefully monitored to ensure that the fits for the calibration intervals converge properly. If not, the fitting procedure is manually adjusted to achieve a precision calibration.

The inconsistencies noted above in the gain between the inner four pixels and the rest of the array for the archived data for this source were due to fitting failures in the time-dependent calibration. These were corrected by the XRISM instrument scientists and XRISM Science Data Center scientists, recovering the energy scale to about 0.5\,eV systematic uncertainty on the central four pixels and better than 0.2\,eV on the remaining pixels at 5.9\,keV. The data were then fully reprocessed using the recovered gain history in advance of the analysis described below.

Finally, using the calibrated data with the improved gain history, we reprocessed the event files with the XRISM pipeline (processing version \texttt{03.00.011.008}). Only High Primary (Hp) grade events were selected, following the standard event-screening criteria described in \citet{Mochizuki2025}. The energy range was restricted to 2--10\,keV, and the spectra were re-binned by combining adjacent detector channels by factors of 6 in the 2--10\,keV energy range and 5 in the 5.5--7.5\,keV range, in order to improve the statistical quality while preserving the spectral resolution required for the analysis. Excluding the low-statistics bins around 2\,keV, the shortest-exposure spectrum contains a minimum of 30 counts per bin and a mean of approximately 60 counts per bin, ensuring adequate statistical quality for the analysis.

\begin{table*}[h!]
\centering
\caption{Orbital phase intervals selected for the spectral analysis of \cenx. The start and stop times are given in MJD (TT).}
\begin{tabular}{l c c c c c}
\hline
Phase & GTI &
{MJD$_{\rm start}$} & {MJD$_{\rm stop}$} &
{Duration} & \boldmath$\phi_{\rm start}$--\boldmath$\phi_{\rm stop}$ \\
\hline
Egress      & End of GTI 3    & 60353.411081 & 60353.455787 & 3.86 ks & 0.10--0.13 \\
Post-egress & GTI 5           & 60353.539352 & 60353.591319 & 4.49 ks & 0.17--0.19 \\
Pre-ingress & GTI 26          & 60354.953588 & 60355.017130 & 5.49 ks & 0.84--0.87 \\
Ingress     & Start of GTI 27 & 60355.058766 & 60355.087741 & 2.50 ks & 0.88--0.90 \\
\hline
\end{tabular}
\label{tab:gti_intervals}
\end{table*}

\subsection{Time Interval Selection}

\begin{figure*}[h!]
    \centering
    \includegraphics[width=\textwidth]{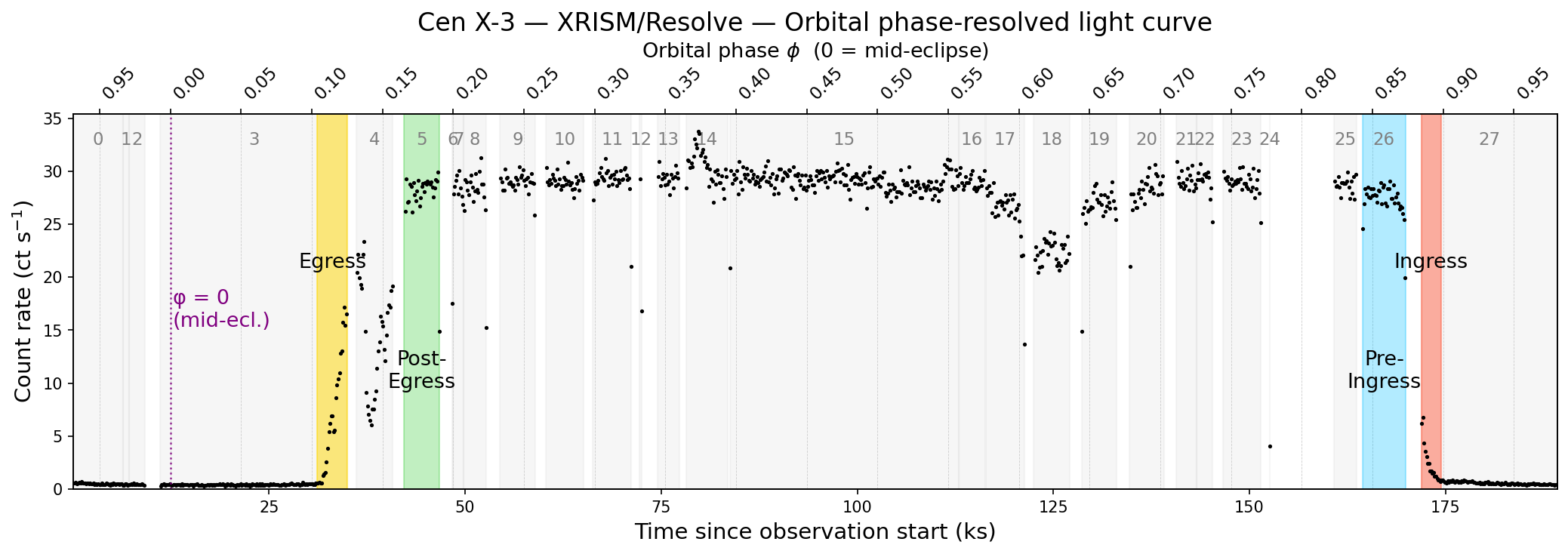}
    \caption{\xrism\ background-subtracted 2--10\,keV light curve of \cenx, binned to 100\,s resolution. The shaded regions mark the time intervals selected for the spectral analysis, including the eclipse egress and ingress intervals, as well as the post-egress and pre-ingress} reference intervals.
    \label{fig:lc}
\end{figure*}

The eclipse egress and ingress phases were identified from the \xrism\
2--10\,keV light curve (Fig.~\ref{fig:lc}) using the Good Time Intervals (GTIs)
defined during the standard data screening. The GTI numbering follows the
default segmentation of the processed event file. The selected egress and
ingress intervals correspond to the rising and declining portions of the
light curve associated with the orbital eclipse transitions of \cenx.

In addition, two out-of-eclipse (OOE) reference intervals were selected to
characterize the unobscured emission of the system, one immediately after
egress (GTI~5), excluding the dipping episode, and one immediately before
ingress (GTI~26).

The complete set of intervals used in the spectral analysis, together with
their corresponding orbital phases, is summarized in Table~\ref{tab:gti_intervals}. Fig.~\ref{fig:lc} shows the full \xrism\
light curve with the numbered GTIs and highlights the intervals adopted in this work.

\begin{figure*}[htpb]
    \centering
    \includegraphics[width=\textwidth]{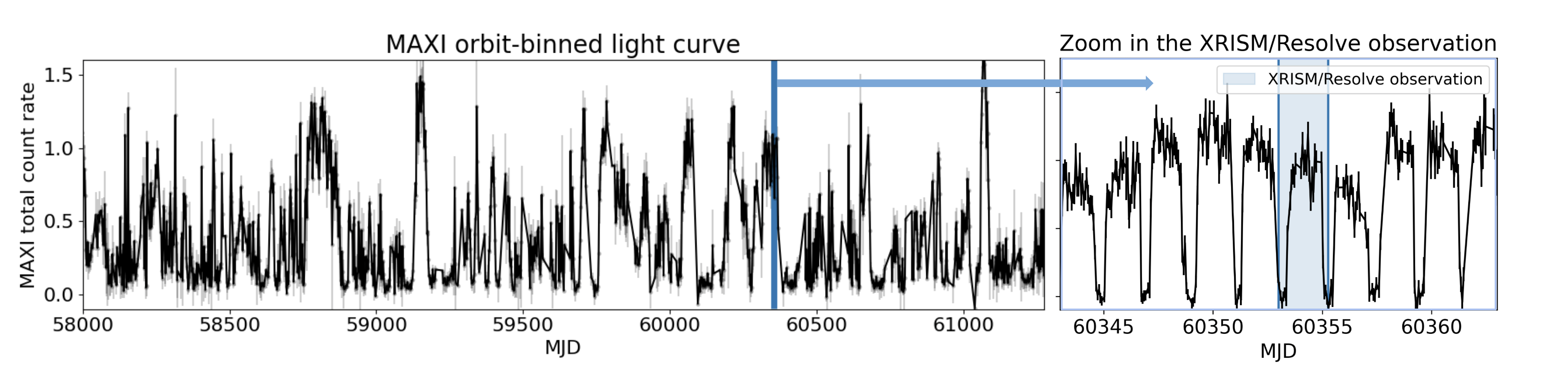}
    \caption{Long-term MAXI light curve of \cenx, using data from
    orbital phases $\phi=0.4$--$0.6$ to minimize eclipse-related
    modulations and trace the intrinsic variability of the source.
    Each point represents the error-weighted mean count rate over a
    complete orbital cycle. The left panel shows the long-term
    evolution of the source, with the blue vertical line marking the
    time of the \xrism\ observation analyzed in this work. The right
    panel shows a zoom-in of the same MAXI light curve around the
    \xrism\ observation; the two blue vertical lines indicate the
    start and stop times of the \xrism\ observation. The source was
    observed during a high-soft state, coinciding with a significant
    increase in X-ray flux.}
    \label{fig:maxi}
\end{figure*}

\section{Results}
\label{sec:results}
\subsection{State of the source during the observation}

In order to characterize the accretion state of \cenx\ during the \xrism\ observation, we examined long-term X-ray monitoring from the \textit{Monitor of All-sky X-ray Image} \citep[MAXI;][]{Matsuoka2009}. The MAXI orbital light curve indicates that the source was in a high-soft state at the time of the observation (Fig.~\ref{fig:maxi}), in agreement with the elevated soft X-ray flux and the overall morphology of the XRISM light curve (Fig.~\ref{fig:lc}). 

As discussed in Sect.~\ref{sec:introduction}, the high-soft state is characterized by a higher X-ray luminosity and a more prominent soft spectral component compared to the low/hard state \citep{2021MNRAS.501.5892S}. In this context, \citet{2010RAA....10.1127D} showed that eclipse ingress and egress transitions in Cen\,X-3 are significantly sharper and shorter during high-soft states, while they become more gradual and extended in low-hard states. This behaviour was interpreted as evidence for variable obscuration associated with a precessing accretion disk.
The eclipse profile observed in the XRISM light curve (Fig.~\ref{fig:lc}) is fully consistent with this scenario.

\subsection{Spectral analysis}
\label{sec:model}

To characterize the continuum emission, local absorption, and fluorescent reprocessing features observed in the XRISM/Resolve spectra, we adopted a physically motivated multi-component model implemented in SPEX 3.08.03 \citep{1996uxsa.conf..411K,https://doi.org/10.5281/zenodo.19694987}. The model includes interstellar and local absorption, a Comptonized continuum, a soft thermal component, a reflection component accounting for fluorescence and Compton scattering, and additional Gaussian components to reproduce emission lines. The model structure is:

\begin{equation}
\begin{split}
    \mathcal{M} =
    \texttt{hot}_\mathrm{ISM} \times
    \Bigl[
        \texttt{hot}_\mathrm{local} \times\bigl(  \texttt{vgau}_\mathrm{edge}
        \times \texttt{pow} \;+\;  \texttt{bb}\bigr)
\\
        \;+\;
        \texttt{xskirtor}
        \times \texttt{vgau}_\mathrm{refl}         \;+\;
        \sum \texttt{gaus}
    \Bigr]
\end{split}
\label{eq:model}
\end{equation}

\noindent where \texttt{hot}$_\mathrm{ISM}$ represents the interstellar  absorption, \texttt{hot}$_\mathrm{local}$ accounts for absorption intrinsic to the binary system, \texttt{pow} and \texttt{bb} describe the Comptonized and thermal continuum components, respectively, and \texttt{xskirtor} models the reflected emission \citep{2023A&A...674A.123V}. The two velocity broadening components, \texttt{vgau}$_\mathrm{edge}$ and \texttt{vgau}$_\mathrm{refl}$, account for the broadening of absorption edges and reflection features, respectively. 

For the Hp events used in this analysis, the Resolve spectral resolution is approximately 4.5\,eV FWHM at 6\,keV \citep{Mochizuki2025}, corresponding to a velocity scale of $\sim225$\,km\,s$^{-1}$ FWHM. This value represents the instrumental resolution rather than a strict lower limit on the intrinsic velocity broadening that can be measured, which also depends on the statistical quality of the spectrum and the complexity of the spectral model. The model components are subsequently convolved with the instrumental redistribution matrix (RMF); therefore, the velocity-broadening parameters reported here correspond to the intrinsic broadening of the spectral components and do not include the instrumental contribution. A separate limit applies to velocity shifts, which are constrained by the absolute energy-scale calibration accuracy of Resolve, recommended at $\pm0.3$\,eV in the 5.4--9\,keV band \citep{10.1117/1.JATIS.11.4.042018}. At the energy of the Fe\,K$\alpha$ line ($\sim$6.4\,keV), this corresponds to an energy-scale uncertainty of $\sim14$\,km\,s$^{-1}$.

Additional Gaussian components were included to model residual emission features in the Fe K band. 

The continuum model — a power law combined with a blackbody soft excess — follows the approach established by previous studies of Cen~X-3 \citep{2000ApJ...530..429B, Naik2011, 2021MNRAS.501.5892S}. No high-energy cutoff component was included, as the cutoff energy reported in those works ($\sim$14\,keV; \citealt{2000ApJ...530..429B}) lies well above the 1.7--12\,keV band of XRISM/Resolve.
The best-fit parameters for all four intervals are listed in Table~\ref{tab:comparison}.
Fig.~\ref{fig:spectra} and Fig.~\ref{fig:fekband} show the
spectra and best-fit models.

\subsubsection{Interstellar Absorption (\texttt{hot}$_\mathrm{ISM}$, fixed)}
\label{sec:hotISM}

The Galactic ISM column was fixed to
$N_\mathrm{H}^\mathrm{ISM}=1.11\times10^{22}$\,cm$^{-2}$
based on the HI4PI 21-cm survey
\citep{2016A&A...594A.116H}, consistent with the optical
reddening toward \vstar\ \citep{2007A&A...473..523V}.
We modeled the ISM absorption using the SPEX \texttt{hot} model
with the temperature fixed at $T=10^{-6}$\,keV, which corresponds
to the minimum temperature adopted by SPEX for this model and
produces an effectively neutral absorber. Thus, although the
component is formally labeled \texttt{hot}, it is used here as a
neutral ISM absorption model rather than to represent physically
hot gas \citep{2004A&A...423...49D, 2005A&A...434..569S}.

\subsubsection{Local absorber (\texttt{hot}$_\mathrm{local}$)}

\label{sec:hotlocal}

To model the cold to mildly ionized material intrinsic to the binary system, we applied a local absorber component (\texttt{hot}$_\mathrm{local}$) to the continuum. The high-resolution XRISM/Resolve spectra reveal a prominent Fe K-shell absorption edge during both the ingress and egress phases, alongside several weaker, newly resolved absorption edges. To account for these features, the abundances of Fe, Ni, S, Ca, and Ar within \texttt{hot}$_\mathrm{local}$ were initially allowed to vary up to twice their solar values in one spectrum and were subsequently fixed to the resulting best-fit values for the remaining spectra, while all remaining elements were fixed to solar abundances \citep{2009LanB...4B..712L}.

A velocity broadening component, \texttt{vgau}$_\mathrm{edge}$, was coupled to
the local absorber to test for possible velocity broadening of the Fe~K
absorption edge. We compared fits with this component free against the null hypothesis in which the intrinsic edge broadening is fixed to zero. For
eclipse ingress, allowing the broadening to vary yields only a marginal
improvement in the fit statistic, while for egress no improvement is found.
We consider this insufficient to claim a robust detection of intrinsic edge
broadening. Moreover, systematic effects not included in the present model
-- such as a partial Fe\,\textsc{ii} contribution from mild ionization, or
intrinsic fine structure of the Fe\,K edge \citep[e.g.,][]{2004ApJ...612..308J,2006ApJ...648.1066J}
-- could plausibly mimic or mask genuine velocity broadening. We therefore do
not report a measured value for $\sigma_\mathrm{edge}$ and do not interpret
this test as evidence of kinematic broadening of the edge.

The local column density exhibits significant variation across the four
orbital intervals. It ranges from
$N_\mathrm{H}^\mathrm{local}=({4.3^{+1.1}_{-0.4}})\times10^{22}\,\mathrm{cm^{-2}}$
during pre-ingress to
$N_\mathrm{H}^\mathrm{local}=({5.95^{+0.07}_{-0.03}})\times10^{23}\,\mathrm{cm^{-2}}$
during ingress, with intermediate values of
$({4.4^{+0.9}_{-1.2}})\times10^{22}\,\mathrm{cm^{-2}}$
during post-egress and $({1.76\pm0.08})\times10^{23}\,\mathrm{cm^{-2}}$
during egress.

The temperature of the local absorber, $t^\mathrm{local}$, is generally
very low. It is estimated to be $({16.3^{+1.0}_{-2.4}})$\,eV
during egress, while only upper limits are obtained for the remaining
phases: $t^\mathrm{local}{\leq126}$\,eV during post-egress,
$t^\mathrm{local}{\leq36}$\,eV during pre-ingress, and
$t^\mathrm{local}{\leq0.81}$\,eV during ingress. These low
temperatures indicate that the local absorber is predominantly neutral or
only weakly ionized.

Because $N_\mathrm{H}^\mathrm{local}$ and $t^\mathrm{local}$ can be correlated in the \texttt{hot} model, we additionally investigated their joint confidence contours (Fig.~\ref{fig:contour_plots}). The contours indicate that $N_\mathrm{H}^\mathrm{local}$ remains constrained despite its correlation with the absorber temperature. For post-egress, pre-ingress, and ingress, the confidence contours do not close on the lower side in $t^\mathrm{local}$, as the allowed range reaches SPEX's minimum temperature for the \texttt{hot} model ($10^{-6}$\,keV); since this lower bound is imposed by the model rather than constrained by the data, we report only the 68\% upper limit for $t^\mathrm{local}$ in these three phases (Table~\ref{tab:comparison}).

The prominent Fe\,K$\alpha$ edge and the spectral structure around $\sim$7\,keV, particularly during the ingress phase, are reminiscent of the dust-absorption features reported by \citet{Costantini_2023} in spectra affected by olivine and dust components. However, in order to avoid overcomplicating the spectral model by introducing an additional dust absorption component beyond the hot absorber already included, this component was not incorporated in the final fits. Nevertheless, its contribution cannot be ruled out, as it was tested and found to provide a statistically acceptable description of the data in exploratory fits.

\subsubsection{Comptonized Continuum (\texttt{pow})}
\label{sec:continuum}

The power-law photon index varies with orbital phase. The spectrum is harder during egress than during the rest of the orbit, whereas ingress is characterized by a softer spectrum, with $\Gamma = 1.13$ and $1.42$, respectively. The OOE measurements remain confined to a narrow range of $\Gamma = 1.48$--$1.52$, indicating limited spectral variability outside the eclipse transitions.

Because the photon index can be correlated with the absorbing
column density, particularly in the presence of strong local absorption, we
also examined the joint confidence contours between $\Gamma$ and
$N_\mathrm{H}^\mathrm{local}$ (Fig.~\ref{fig:contour_plots}). The contours show only weak covariance between the two parameters for most orbital intervals, although a degree of degeneracy is present during pre-ingress.
Therefore, the contribution of the
$N_\mathrm{H}^\mathrm{local}$--$\Gamma$ degeneracy cannot be completely
excluded, particularly during pre-ingress.

\subsubsection{Blackbody (\texttt{bb})}
\label{sec:bb}

The blackbody component is required in the egress, post-egress, and pre-ingress spectra, while its normalization becomes consistent with zero within the uncertainties during ingress. However, the blackbody normalization shows a strong degeneracy with the local absorbing column density, as illustrated by the joint confidence contours between $N_\mathrm{H}^\mathrm{local}$ and the blackbody normalization (Fig.~\ref{fig:contour_plots}). We therefore cannot reliably attribute the apparent orbital variation of the blackbody normalization to intrinsic variability of the soft component. The absence of a significant blackbody contribution during ingress should also be interpreted with caution, since this spectrum has the lowest number of counts and the strongest local absorption.

Furthermore, XRISM/Resolve has limited sensitivity below $\sim$2\,keV, where the blackbody contributes most strongly. Consequently, its temperature and normalization are only weakly constrained and are particularly sensitive to degeneracies with the absorption and continuum parameters.

\subsubsection{Disk Reflector (\texttt{xskirtor})}
\label{sec:reflector}

The \texttt{xskirtor} component models X-ray reprocessing by cold material, including Compton scattering and fluorescent Fe K$\alpha$ and Fe K$\beta$ emission \citep{2023A&A...674A.123V}. 

Across all four temporal intervals, the structural parameters of the reflector were kept frozen. The intrinsic column density and covering fraction of the reflector reached their upper bounds at $N_\mathrm{H} = 10^{25}$\,cm$^{-2}$ and $f_\mathrm{cov} = 0.95$, respectively. The inclination parameter was held constant at $\cos i = 0.24$ ($i \approx 76^\circ$), which is in the inclination range derived in \cite{2021MNRAS.501.5892S}.

The reflection normalization exhibits variability, decreasing from $380$ during egress to $50$ during ingress, and reaching $87$ in the post-egress (GTI 5) and pre-ingress (GTI 26) intervals. 

\subsubsection{Ionized emission lines}
\label{sec:feklines}

\begin{figure*}[h!]
    \centering
    \includegraphics[width=\textwidth]{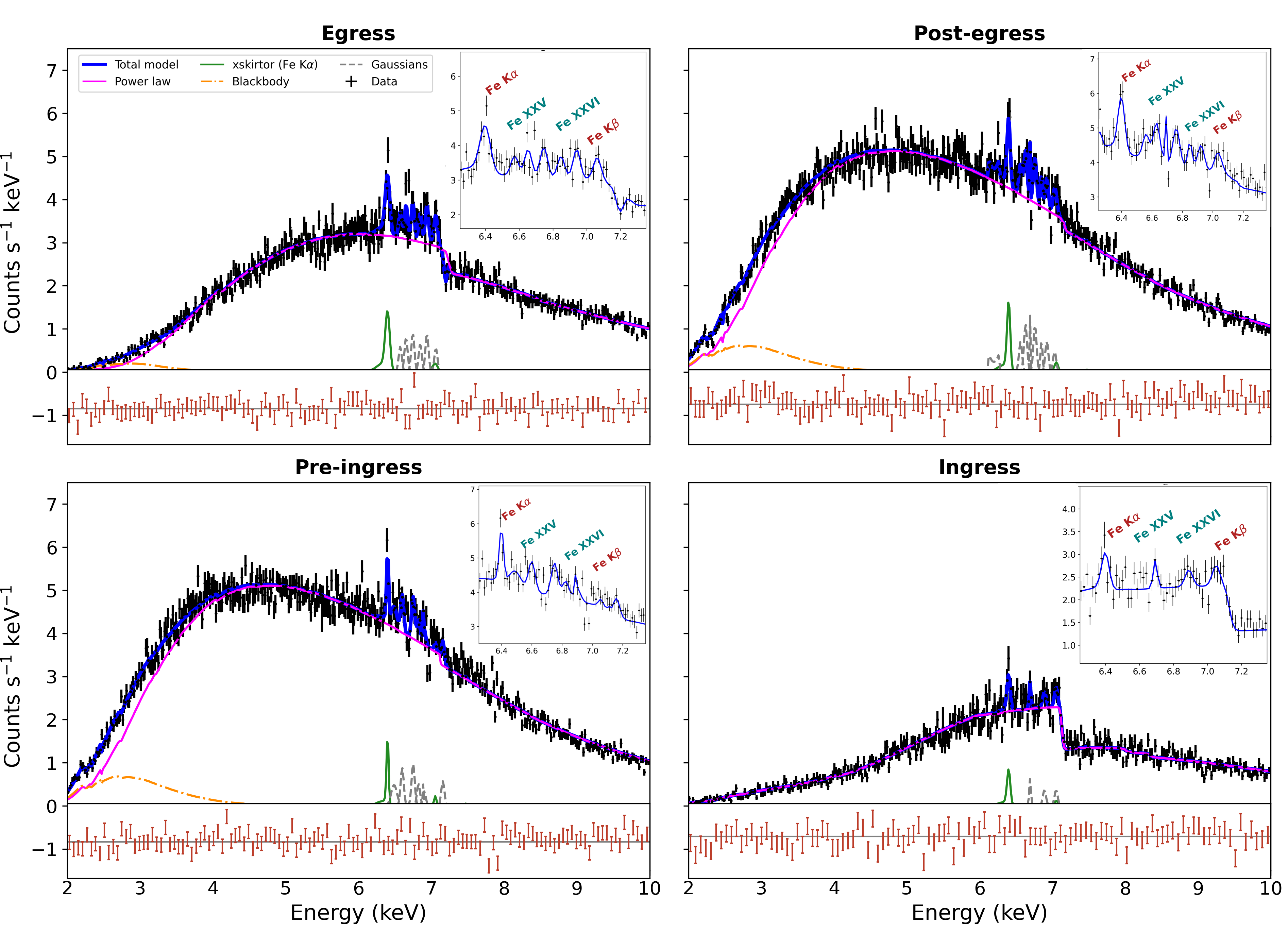}
    \caption{\xrism\ spectra of \cenx\ during the egress, out-of-eclipse (OOE), and ingress orbital intervals. Black points show the observed data, while the blue solid line represents the total best-fit model. The individual spectral components are also displayed: the pink solid line corresponds to the Comptonized power-law continuum, the orange dashed line to the blackbody thermal emission, the green solid line to the reflected Fe\,K$\alpha$ component modeled with \texttt{xskirtor}, and the gray dashed lines to the highly ionized emission lines. The lower panel of each spectrum displays the residuals with respect to the best-fit model. For each spectrum, the upper-left inset shows a zoomed-in view of the 6.25--7.25\,keV Fe\,K band.}
    \label{fig:spectra}
     \label{fig:fekband}
\end{figure*}

The OOE and egress phases—where a continuum excess is observed in the Fe K band—are characterized by a complex Fe emission line structure. This profile is modeled using Gaussian components to resolve the iron K band complex and related transitions, such as Fe \textsc{xxv} and Fe \textsc{xxvi}. However, due to a lower signal-to-noise ratio this detailed line emission cannot be resolved during the ingress phase. While these additional Gaussians were introduced to improve the fit statistics and are shown in Fig. \ref{fig:fekband}, a detailed discussion of these lines is beyond the scope of this work.

\begin{table*}[t]
\centering
\caption{Comparison of the best-fit spectral parameters for
  the four \cenx\ orbital intervals.}
\label{tab:comparison}
\resizebox{\textwidth}{!}{
\begin{tabular}{llcccc}
\toprule
Component & Parameter &
  Egress & post-egress (GTI\,5) &
  pre-ingress (GTI\,26) & Ingress \\
\midrule

\texttt{hot}$_\mathrm{ISM}$ (fixed)
  & $N_\mathrm{H}^\mathrm{ISM}$ ($10^{22}$\,cm$^{-2}$)
  & \multicolumn{4}{c}{1.11 (frozen)} \\
  & $t^\mathrm{ISM}$ (keV)
  & \multicolumn{4}{c}{$1\times 10^{-6}$ (frozen)} \\
  \\

\hline
\\
\texttt{hot}$_\mathrm{local}$
  & $N_\mathrm{H}^\mathrm{local}$ ($10^{22}$\,cm$^{-2}$)
  & $17.6\pm0.8$
  & $4.4_{-1.2}^{+0.9}$
  & $4.3_{-0.4}^{+1.1}$
  & $59.5^{+0.7}_{-0.3}$ \\

  & $f_\mathrm{cov}$
  & $1.0^{+0.0}_{-1.0}{}^{a}$
  & $1.0^{+0.0}_{-1.0}{}^{a}$
  & $1.00_{-0.6}^{+0.0}$
  & $0.942_{-0.002}^{+0.007}$ \\

  & $t^\mathrm{local}$ (eV)
  & $16.3_{-2.4}^{+1.0}$
  & $\leq 126$$^{c}$
  & $\leq 36$$^{c}$
  & $\leq 0.81$$^{c}$ \\
\\
\hline
\\
\multirow{3}{*}{\texttt{pow}}
  & Norm ($10^{48}$\,ph\,s$^{-1}$\,keV$^{-1}$)
  & $4.5_{-0.1}^{+0.4}$
  & $11.35_{-0.08}^{+0.02}$
  & $11.3_{-0.4}^{+0.5}$
  & $7.6_{-1.4}^{+2.0}$ \\

  & $\Gamma$
  & $1.13\pm{0.04}$
  & $1.483_{-0.001}^{+0.004}$
  & $1.52_{-0.03}^{+0.01}$
  & $1.42_{-0.11}^{+0.09}$ \\

  & Flux$_{2-10}^{b}$
  & $17_{-0.5}^{+1.5}$
  & $29.49_{-0.20}^{+0.04}$
  & $28.2_{-0.9}^{+1.2}$
  & $7.1_{-1.4}^{+1.8}$ \\
\\
\hline
\\
\multirow{3}{*}{\texttt{bb}}
  & Norm ($10^{16}$\,m$^2$)
  & $12_{-1}^{+3}$
  & $<8.6\times10^{-2}$
  & $\left(10_{-3}^{+2}\right)\times10^{-2}$
  & $-$ \\

  & $kT$ (eV)
  & $211_{-6}^{+4}$
  & $314_{-4}^{+1}$
  & $310_{-10}^{+90}$
  & $-$ \\

  & Flux$_{2-10}^{b}$
  & $3.5_{-0.2}^{+0.9}$
  & $<22.4$
  & $1.9\pm{0.5}$
  & $-$ \\
\\
\hline
\\
\texttt{xskirtor}
  & Norm (Global)
  & $380_{-50}^{+60}$
  & $87_{-12}^{+13}$
  & $87_{-11}^{+12}$
  & $50_{-10}^{+11}$ \\

  & Flux$_{2-10}^{b}$
  & $0.125_{-0.016}^{+0.020}$
  & $1.8_{-0.2}^{+0.3}$
  & $0.060\pm{0.008}$
  & $0.030\pm{0.006}$ \\
\\
\hline
\\
\texttt{vgau}$_\mathrm{refl}$
  & $\sigma_\mathrm{refl}$ (km\,s$^{-1}$)
  & $1400_{-200}^{+300}$
  & $900_{-200}^{+300}$
  & $510_{-130}^{+160}$
  & $900_{-200}^{+300}$ \\
\\
\hline
\hline
\\
  & $C$-stat
  & 466 & 337.6 & 596.4 & 373 \\

  & $C$-stat (exp.)
  & $420 \pm 30$
  & $440 \pm 30$
  & $450 \pm 30$
  & $380 \pm 30$ \\

  & d.o.f.
  & 388 & 433 & 423 & 366 \\
\\
\bottomrule
\end{tabular}
}
\tablefoot{
$^{a}$ Uncertainties that are not well constrained.
$^{b}$ Fluxes in the 2--10\,keV band, in units of
$10^{-14}$\,W\,m$^{-2}$.
$^{c}$ 68\% upper limit; the confidence contour does not close on
the lower side, as the allowed range runs into SPEX's minimum
temperature for the \texttt{hot} model ($10^{-6}$\,keV).
}
\end{table*}

\subsection{Fe\,K$\alpha$ Time Delay Analysis}
\label{sec:ccf}

We searched for a time delay between the continuum (3--6\,keV)
and the \feka\ (6.2--6.6\,keV) light curves during both the
egress (GTI\,3, $\Delta t\approx3.86$\,ks, $\sim$38 bins) and
ingress (GTI\,27, $\Delta t\approx2.5$\,ks, $\sim$25 bins)
intervals. We used the Cross-Correlation Function (CCF) with
100\,s time bins and Monte Carlo uncertainties derived from
$N=1000$ realizations in which Poisson noise was added to the
\feka\ light curve \citep{1988ApJ...333..646E, 1998PASP..110..660P}.

No significant delay was detected in either phase:
\begin{align}
    \tau_\mathrm{egress}  &= 0 \pm 19\ \mathrm{s}
        \quad (1\sigma,\ \mathrm{MC}), \\
    \tau_\mathrm{ingress} &= 0 \pm 98\ \mathrm{s}
        \quad (1\sigma,\ \mathrm{MC}).
\end{align}

The tighter egress constraint places an upper limit on the distance of the reprocessing material from the NS:
\begin{equation}
    r < c\,(\tau + \sigma_\tau)
      \approx 5.6\times10^{11}\ \mathrm{cm}
      \approx 8 R_\odot,
    \label{eq:rlimit}
\end{equation}
where $R_\odot$ is the solar radius, consistent with \feka\ production in the inner accretion disk
or at the base of the stellar wind, well within the binary
separation of \cenx, $a\sim2\times10^{12}$\,cm;
\citep{1999MNRAS.307..357A}. The larger ingress uncertainty
($\sigma_\tau=98$\,s, $r<2.9\times10^{12}$\,cm) reflects
the lower count rate and fewer time bins ($\sim$25 vs
$\sim$38), and is comparable to the binary separation,
providing no additional geometric constraint. In any case, the Fe K$\alpha$ emission likely arises from multiple regions, a scenario that is discussed in Sect.~\ref{sec:discussion}.

\section{Discussion}
\label{sec:discussion}

\begin{figure}[htpb]
    \centering
    \includegraphics[width=0.5\textwidth]{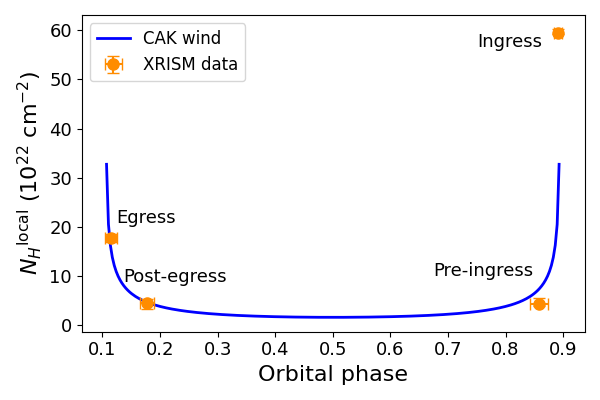}
    \caption{Comparison between the theoretical CAK spherical wind model (blue line) and the local column density (\(N_{\mathrm{H}}^{\mathrm{l}ocal}\)) derived from the four XRISM observations (orange points) across different orbital phases of \cenx.}
    \label{fig:nhmodel}
\end{figure}

\begin{figure}[htbp]
\centering
    \includegraphics[width=0.5\textwidth]{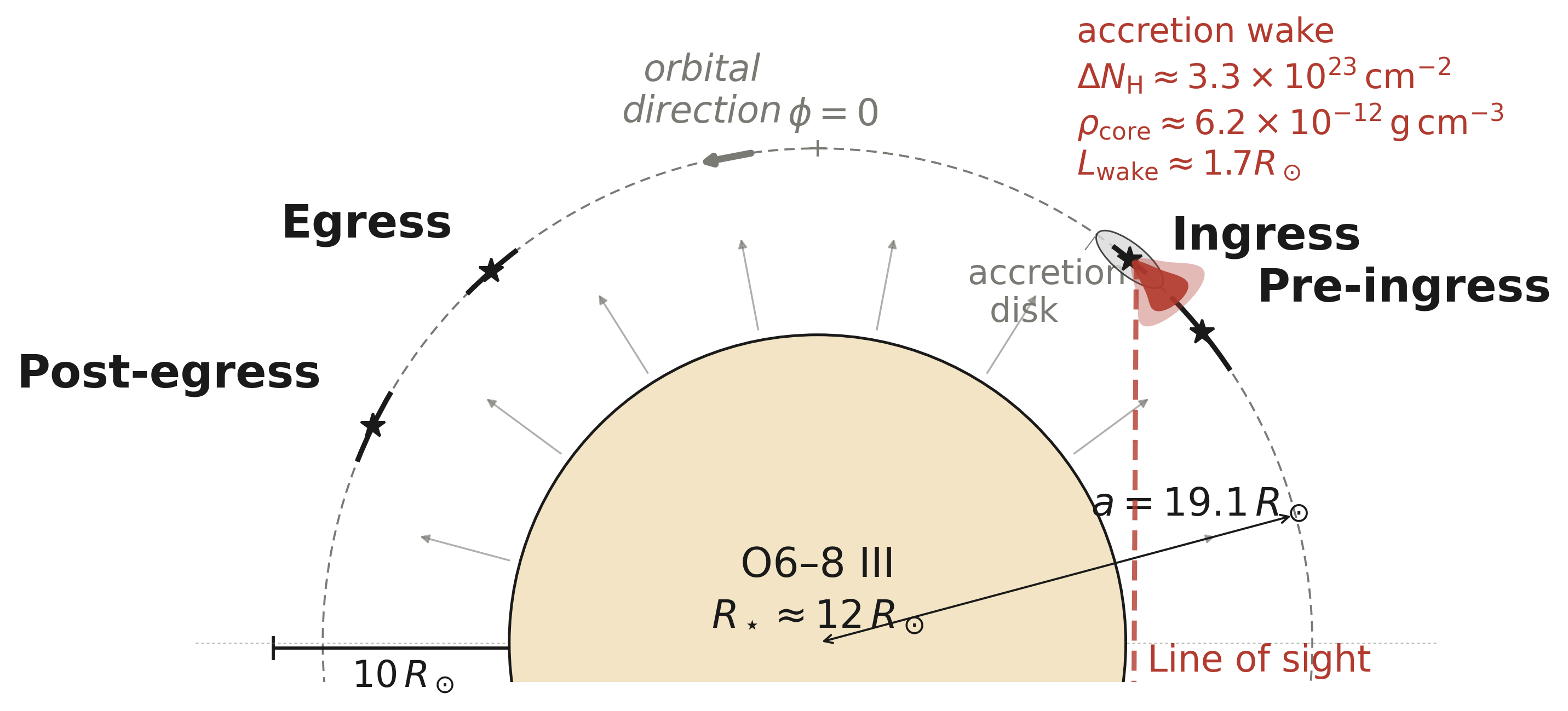}
\caption{Schematic (upper half, to scale in $R_\odot$) of the Cen\,X-3 system near eclipse.  Thick arcs on the orbit ($a = 19.1\,R_\odot$) mark the pre-ingress, ingress, egress, and post-egress phase intervals. The red band between pre-ingress and ingress represents the compact absorbing structure inferred from the $N_{\rm H}$ excess.}
\label{fig:schematic_cenx3}
\end{figure}

The high-resolution spectroscopy provided by XRISM/Resolve has enabled a characterization of the absorbing column, dynamical structure, and
Fe\,K emission complex of \cenx\ across four distinct orbital intervals
covering eclipse egress, eclipse ingress, and two out-of-eclipse (OOE)
reference phases. The spectral resolution of Resolve
(${\sim}5$\,eV FWHM) reveals absorption-edge structures during the eclipse
transitions and disentangles the neutral fluorescent, Compton-scattered, and
highly ionized Fe emission components that are blended in CCD-resolution
spectra.

The main quantitative results are: (i) a nearly two-order-of-magnitude variation in the local absorbing column density, from $N_\mathrm{H}^\mathrm{local}=4.25\times10^{22}$ to $5.9\times10^{23}$\,cm$^{-2}$, with the excess absorption during ingress requiring a localized overdense structure beyond a smooth stellar-wind model; (ii) orbital variation in the Fe\,K$\alpha$ velocity width, ranging from ${\sim}510$ to ${\sim}1400$\,km\,s$^{-1}$, with the broadest profile during egress and statistically consistent widths of ${\sim}900$\,km\,s$^{-1}$ during ingress and post-egress; (iii) orbital variability of the Fe\,K$\alpha$ line flux, with no clear correlation with its velocity width; and (iv) the detection of highly ionized Fe\,\textsc{xxv} and Fe\,\textsc{xxvi} emission throughout the orbit. We discuss each of these results in turn.

\begin{SCfigure*}
\centering
\includegraphics[width=0.85\textwidth]{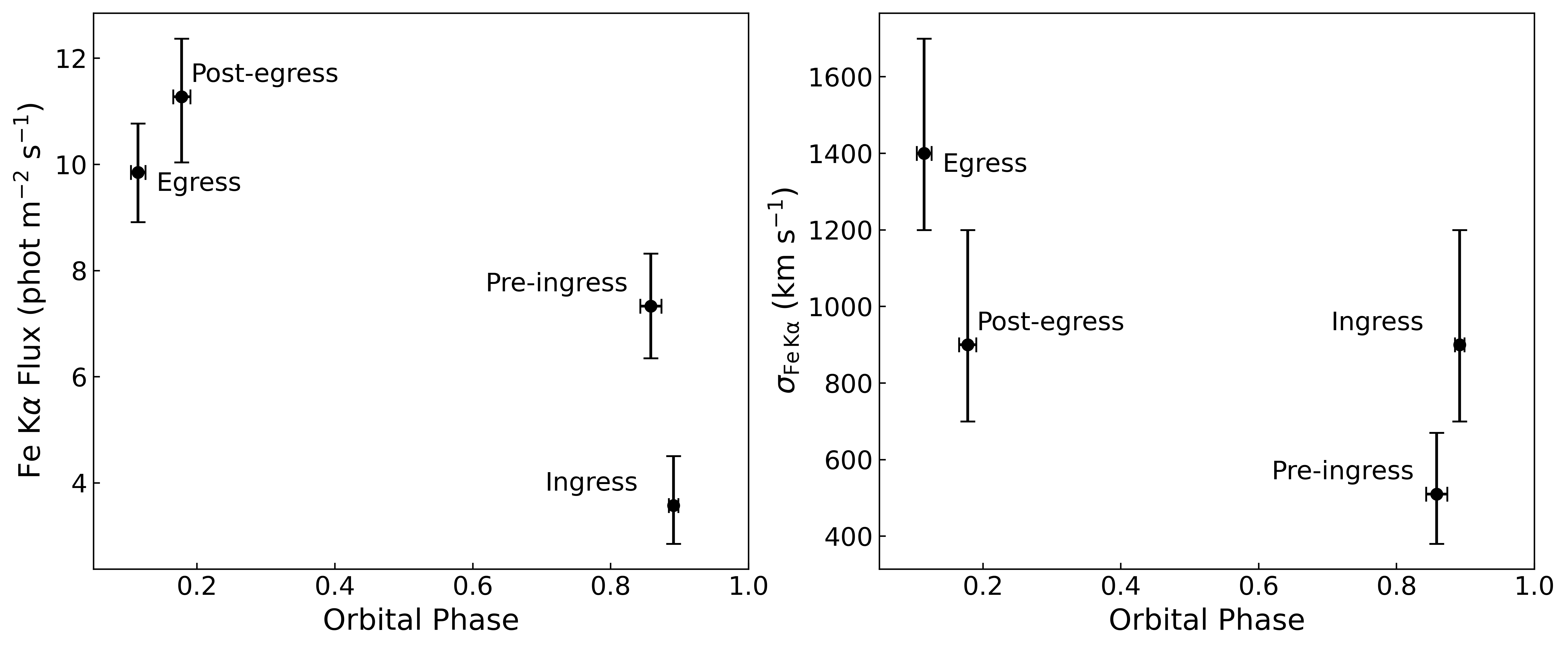}
\caption{Left: Orbital variation of the Fe K$\alpha$ flux $\sigma_{\rm Fe\,K\alpha}$. Right: Orbital variation of the velocity broadening of the Fe\,K$\alpha$ emission line
($\sigma_{\rm Fe\,K\alpha}$) across the four spectral intervals
studied in this work.  Error bars correspond to 90\% confidence intervals.}
\label{fig:fekorbitalphase}
\end{SCfigure*}

\begin{SCfigure*}
\centering
\includegraphics[width=0.84\textwidth]{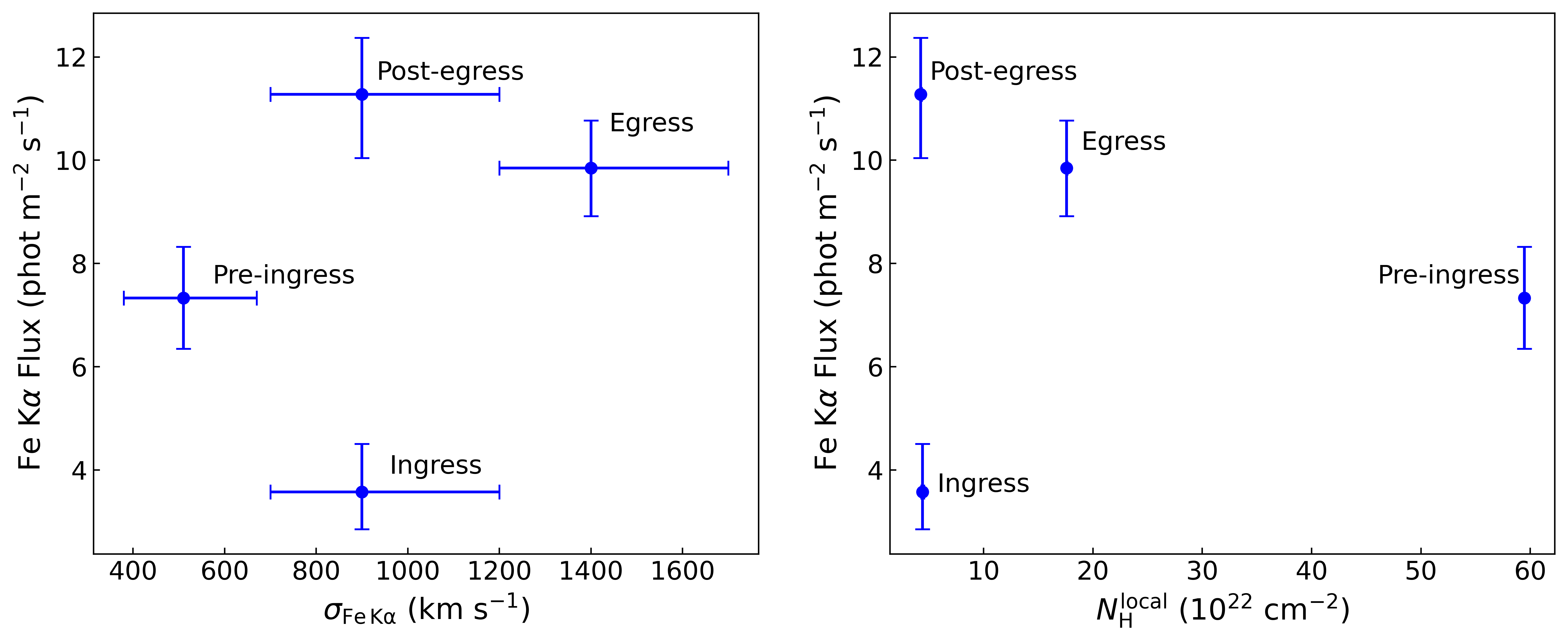}
\caption{
Left: Fe K$\alpha$ flux as a function of the Fe K$\alpha$ line broadening ($\sigma_{\rm Fe\,K\alpha}$). Right: Fe K$\alpha$ flux as a function of the local equivalent hydrogen column density
$N_{\rm H}^{\rm local}$. Error bars correspond to 90\% confidence intervals.
}
\label{fig:fekflux}
\end{SCfigure*}

\subsection{Orbital evolution of the absorbing column}
\label{sec:disc_nh}

\subsubsection{Ingress--egress asymmetry and the smooth-wind baseline}

The local equivalent hydrogen column density varies by nearly two orders of magnitude across the orbit, reaching its maximum during eclipse ingress (Fig.~\ref{fig:nhmodel}). Similar orbital asymmetries have been reported in other eclipsing HMXBs. For example, \citet{2021A&A...647A..75F} measured a column-density enhancement of $\gtrsim3$ in IGR\,J18027$-$2016, while \citet{2015A&A...577A.130F} reported a comparable increase in XTE\,J1855$-$026. 
Within Cen X-3 itself, \cite{Suchy2008} reported enhanced absorption near eclipse ingress and egress in RXTE observations covering two consecutive orbits. They found significant orbit-to-orbit differences in the evolution of the absorbing column density, with a stronger increase in $N_H$ during the second orbit, which they interpreted as evidence for variability in the circumstellar absorbing material, possibly related to an accretion wake. 

Motivated by these results, we investigated whether the orbital evolution of $N_{\rm H}$ measured with XRISM can be reproduced by a smooth stellar wind. For this purpose, we computed the expected column density as a function of orbital phase using a spherically symmetric Castor, Abbott, Klein (CAK) wind model \citep{Castor1975}, in which the wind velocity follows the standard $\beta$-law,

\begin{equation}
v(r)=v_\infty\left(1-\frac{R_\star}{r}\right)^\beta,
\end{equation}

while the density profile is derived from mass conservation,

\begin{equation}
\rho(r)=\frac{\dot{M}}{4\pi r^2 v(r)}.
\end{equation}

To convert the total mass density $\rho(r)$ into hydrogen number density $n_{\rm H}(r)$, we use the hydrogen mass fraction $X_{\rm H} = 0.70$ and the hydrogen atom mass $m_{\rm H}$:
\begin{equation}
     n_{\rm H}(r) = \frac{X_{\rm H} \cdot \rho(r)}{m_{\rm H}}.
\end{equation}

The hydrogen column density is then obtained by integrating the density along the line of sight (LOS) from the NS to the observer,

\begin{equation}
N_{\rm H}(\phi)=\int_{\rm LOS} n_{\rm H}(r)\,ds.
\end{equation}

This calculation provides a smooth-wind baseline, implicitly assuming that the density at a given radius is described by the mean spherically symmetric wind profile. In reality, stellar winds are expected to be structured and clumpy, and density inhomogeneities along the line of sight can introduce additional variations in the measured column density. Such clumping is not included in the present model and therefore represents an additional source of systematic uncertainty when comparing the predicted and observed $N_{\rm H}$ values.

The predicted orbital evolution is shown in Fig.~\ref{fig:nhmodel}. The smooth-wind model reproduces the overall absorption measured during the egress and post-egress phases, indicating that the ambient stellar wind is reasonably described by a standard CAK profile in this part of the orbit. However, the model deviates from the data before entering eclipse, where during the ingress phase, the observed absorption rises sharply relative to the pre-ingress.

To quantify this discrepancy, we fitted the observed $N_{\rm H}$ values using the particle-swarm optimization routine implemented in the \texttt{xraybinaryorbit} package \citep{SanjurjoFerrn2024}. Since the column density is proportional to the stellar mass-loss rate, $\dot{M}$ can be derived as a free parameter. 

When all orbital phases are included, the best-fitting mass-loss rate does 
not explain all the $N_\mathrm{H}$ values observed. Restricting the fit to the OOE and egress phases yields a mass-loss rate of $\dot{M} = \left(7.13 \pm 1.37\right) \times 10^{-7}\,M_\odot\,\mathrm{yr}^{-1}$. This estimate is broadly consistent with previous observational and theoretical determinations for Cen X-3. It is of the same order of magnitude as the theoretical prediction of \(\dot{M} = 5.3 \times 10^{-7}\,M_\odot\,\mathrm{yr}^{-1}\) from \cite{2015A&A...577A.130F} and the observational estimate of \(\dot{M} = 7.8 \times 10^{-7}\,M_\odot\,\mathrm{yr}^{-1}\) obtained by \cite{Klawin2023}.

Conversely, if the ingress phase $N_\mathrm{H}$ value were produced by a 
smooth stellar wind, fitting this phase alone would require a mass-loss rate  of the order of $\dot{M} \sim 5.0 \times 10^{-6} \, M_\odot \, \mathrm{yr}^{-1}$, which is an order of magnitude larger.

This increase in the inferred mass-loss rate therefore indicates that the ingress absorption cannot be explained by a smooth, spherically symmetric wind alone, suggesting the presence of additional dense material along the LOS during ingress. Hydrodynamic simulations predict the formation of accretion wakes and photoionization wakes trailing the NS as a consequence of gravitational focusing and X-ray photoionization of the stellar wind \citep{1990ApJ...356..591B,Blondin1991}. As the system approaches eclipse, the observer's line of sight intersects this dense structure, producing the strong increase in $N_{\rm H}$ observed during ingress. 

Moreover, as shown in Fig.~\ref{fig:nhmodel} the absorption increases sharply from the pre-ingress to the ingress phase. This behavior is consistent with the rapid pre-ingress rise in absorption reported by \citet{refId0} during the high state of Cen\,X-3 (see its Fig. 9).

\subsubsection{Disentangling the nature of the absorbing structure}

\citet{refId0} investigated the orbital evolution of the absorbing column density in Cen\,X-3 MAXI/GSC data and identified two distinct regimes. In the low-intensity state, they reported a smooth increase in $N_{\rm H}$ starting at $\phi_{\rm orb}\sim0.5$, which persists up to eclipse ingress. This behaviour was interpreted in terms of a large-scale accretion and/or photoionization wake extending over a significant fraction of the orbit, and was successfully reproduced using a phenomenological toy model.

In addition, they found evidence for a different absorption pattern during the high-intensity state, characterized by a more abrupt increase in $N_{\rm H}$ close to eclipse ingress (see their Fig.~9), although this component was not modeled in detail.

Our XRISM observation, obtained during a high-intensity state, exhibits a similarly sharp increase in $N_{\rm H}$ confined to a narrow orbital phase interval immediately preceding eclipse ingress. The high spectral resolution of XRISM allows this transition to be clearly resolved, demonstrating that the absorption is not smoothly distributed over the orbit but instead concentrated in a restricted phase range. This behaviour indicates that, in the high state, the dominant absorber is likely associated with a compact structure rather than the extended wake inferred in the low-intensity state of \citet{refId0}. The origin of this transient, localized absorber remains unclear and is not addressed by current phenomenological models.

An alternative explanation for the enhanced absorption observed at eclipse ingress is the presence of dense structures associated with the accretion disk. 
Unlike most wind-fed HMXBs, Cen X-3 is believed to be primarily powered by Roche-lobe overflow, favoring the formation of a persistent accretion disk around the NS \citep{Petterson1978,1986A&A...154...77T,2021MNRAS.501.5892S}. In this scenario, the line of sight may intercept the outer disk rim or the stream--disk impact region as the NS approaches eclipse, producing a rapid increase in $N_{\rm H}$ over a narrow orbital phase interval.
This picture is consistent with \citet{2022RMxAA..58..355T}, who attributed
flux variations during eclipse-egress in Cen\,X-3 to local absorption from
an emerging accretion stream, possibly corotating with the compact object,
and to instabilities at the inner edge of the disk interacting with the
NS magnetosphere.
The localized nature of the observed absorption enhancement is therefore consistent with occultation by disk-related structures, providing an alternative to an extended accretion or photoionization wake. A schematic view of the proposed absorber geometry is shown in Fig.~\ref{fig:schematic_cenx3}.

\subsection{Probing the companion wind: absorption edges and turbulence}
\label{sec:disc_wind}

The NS in Cen\,X-3 acts as an X-ray backlighter illuminating the dense wind
of the O6--8\,III companion. In the XRISM/Resolve spectra this manifests as
resolved K-shell absorption edges from Fe, Ni, S, Ca, and Ar, most prominent
during the eclipse transitions.

\subsubsection{Constraints on the Absorbing Structure} 

To estimate the characteristic properties of the absorbing structure, we
isolate the column density excess relative to the smooth-wind prediction,
$\Delta N_{\rm H} = (32.9^{+0.7}_{-0.3}) \times 10^{22}\,\mathrm{cm}^{-2}$,
and use the orbital phase interval over which it persists to constrain its
geometry. The azimuthal extent of the ingress excess spans
$\Delta\phi_{\rm ing} \approx 0.014$ in orbital phase, or
$\Delta\theta \approx 5^{\circ}$ along the orbital arc, yielding a maximal
projected path length of
\begin{equation}
L_{\rm wake} = 2\pi a\,\Delta\phi_{\rm ing} \approx 1.7\,R_\odot,
\end{equation}
where $a = 19.1\,R_\odot$ is the orbital separation. 

Assuming this excess is distributed homogeneously along $L_{\rm wake}$, the mean volumetric mass density of the absorbing structure is
\begin{equation}
\rho_{\rm wake} =
\frac{\Delta N_{\rm H}\,\mu\,m_{\rm H}}{L_{\rm wake}}
\approx 6.2\times10^{-12}\,\mathrm{g\,cm}^{-3},
\end{equation}
for $\mu=1.3$, corresponding to a hydrogen number density
$n_{\rm H} \approx 2.8\times10^{12}\,\mathrm{cm}^{-3}$.

This exceeds the local smooth-wind density by a factor of ${\sim}300$, confirming that the ingress absorber is a dense, compact structure rather than a modest overdensity in the ambient wind. We note that this value is nearly an order of magnitude higher than the boundary density of $\rho_\mathrm{wake} \approx 3\times10^{-13}\,\mathrm{g\,cm}^{-3}$ adopted in the toy model of \citet{refId0} for the accretion/photoionization wake, where the density was assumed to decrease to zero toward the interior of the wake. This contrast supports the interpretation that the ingress excess in our high-state observation arises from a more compact and denser structure than the
extended, low-density wake invoked to explain the gradual $N_{\rm H}$ rise in the low-intensity state. 

This estimate should be regarded as an order-of-magnitude characterization, since it assumes that the column-density excess is homogeneously distributed along the line of sight. In particular, any clumpiness or substructure in the absorbing material would introduce an additional systematic uncertainty in the inferred density.

A schematic illustration of Cen X-3 and the compact absorbing structure is
shown in Fig.~\ref{fig:schematic_cenx3}. Its narrow azimuthal extent suggests
that the feature originates from a geometrically confined region in the
immediate vicinity of the NS. This may be consistent with emission or
absorption occurring in dense structures associated with the accretion
environment, such as the outer regions of the disk or the interface between
the disk and the stellar wind, although a unique geometric identification is
not possible from the present data alone.

This interpretation is consistent with long-term studies of Cen\,X-3, which have shown that its flux states and orbital variability can be influenced by complex and time-variable obscuration, commonly attributed to a structured accretion disk and its associated absorbing environment \citep{Raichur2008}.

\subsection{Iron K-shell emission lines}

\subsubsection{Fe\,K$\alpha$ velocity broadening and the reprocessing geometry}

The intrinsic velocity width of the Fe\,K$\alpha$ fluorescence line varies across the orbital phases (Fig.~\ref{fig:fekorbitalphase}, right). The broadest profile is measured during egress, with $\sigma_{\rm Fe\,K\alpha}=1400^{+300}_{-200}$\,km\,s$^{-1}$,
while a lower but consistent width of
$900^{+300}_{-200}$\,km\,s$^{-1}$ is measured during
ingress. Before eclipse, the line is narrower, with
$\sigma_{\rm Fe\,K\alpha}=510^{+160}_{-130}$\,km\,s$^{-1}$ during
pre-ingress, while post-egress gives
$\sigma_{\rm Fe\,K\alpha}=900^{+300}_{-200}$\,km\,s$^{-1}$. This results may indicate that the Fe\,K$\alpha$ emission does not originate from a single homogeneous reprocessing region. 

The narrowest measured width (pre-ingress,
$\sigma_{\rm Fe\,K\alpha}=510^{+160}_{-130}$\,km\,s$^{-1}$) lies above the
instrumental velocity scale of $\sim225$\,km\,s$^{-1}$ FWHM discussed in
Sect.~\ref{sec:model}. The ability to measure intrinsic broadening also
depends on the statistical quality of the data; in the 5.5--7.5\,keV band,
the minimum S/N is $\sim7$ per bin. The ingress and post-egress widths are
statistically consistent within their uncertainties, while the egress width
is only marginally larger. The pre-ingress measurement has the lowest
central value.

Fluorescent Fe\,K$\alpha$ emission is expected to arise from several reprocessing sites in HMXBs, including the stellar wind, the accretion stream, the surface of the accretion disk, and other dense circumstellar structures surrounding the NS \citep[e.g.,][]{2003ApJ...582..959W,Torrejn2010,https://doi.org/10.48550/arxiv.1501.03636,2021MNRAS.501.5892S}. During eclipse ingress and egress the line of sight is expected to cross the most dynamically complex regions of the system, where these structures overlap in projection. The observed broadening in these phases, specially in the egress phase, may therefore reflect the superposition of emission from gas with different projected velocities, rather than a single kinematic component.

In contrast, the much narrower Fe\,K$\alpha$ line measured during
pre-ingress is unlikely to originate in the same environment. Assuming Keplerian rotation around a $1.34\,M_\odot$ NS, the
$\sigma_{\rm Fe\,K\alpha}\approx510$\,km\,s$^{-1}$ velocity dispersion measured pre-ingress corresponds to a reprocessing radius roughly a factor of $\sim7$ larger than that implied by the broad $\sigma_{\rm Fe\,K\alpha}\approx1400$\,km\,s$^{-1}$ component observed during egress. This favors an origin in more slowly moving circumstellar material, such as the stellar wind or the environment surrounding the optical companion. 

Taken together, these results suggest that the Fe\,K$\alpha$
fluorescence in Cen\,X-3 is produced by multiple reprocessing regions. A broad component dominates during eclipse egress, ingress, and post-egress, when the line of sight probes the dense accretion environment surrounding the NS, whereas a narrower component becomes more prominent in the pre-ingress, consistent with fluorescence from the extended circumstellar material associated with the donor star and its stellar wind.

\subsubsection{Variation of the line flux} 

No clear correlation is found between the Fe\,K$\alpha$ line flux and its velocity width (Fig.~\ref{fig:fekflux}, left). The ingress and post-egress intervals, for instance, show the same velocity width, $\sigma_{\rm Fe\,K\alpha}=900^{+300}_{-200}$ km s$^{-1}$, but markedly different fluorescence fluxes of $3.58^{+0.93}_{-0.72}$ and $11.28^{+1.10}_{-1.24}$ photons m$^{-2}$ s$^{-1}$, respectively. The post-egress flux is therefore $\sim3.2$ times higher than during ingress, despite their comparable line widths. This suggests that the line flux and velocity width are not directly coupled, with the processes governing the line intensity potentially being decoupled from those responsible for its kinematic broadening, in agreement with the results reported by \citet{Mochizuki_2024}.

The Fe\,K$\alpha$ flux also varies systematically with orbital phase (Fig.~\ref{fig:fekorbitalphase}, left), being higher during egress and post-egress, as the NS emerges from eclipse, than during ingress and pre-ingress, as it approaches eclipse. 
In particular, the strong suppression of the Fe\,K$\alpha$ flux during ingress coincides with the highest local absorbing column, $N_{\rm H}^{\rm local}\approx59\times10^{22}$\,cm$^{-2}$ (Fig.~\ref{fig:fekflux}, right). The persistence of the Fe\,K$\alpha$ suggests that an additional, more extended fluorescent component remains visible throughout the orbital cycle.

\subsubsection{Ionization structure of the Fe\,K complex}

We detect emission from highly ionized Fe\,\textsc{xxv} and Fe\,\textsc{xxvi} in all orbital phases (see Fig.~\ref{fig:fekband}). Several components of the Fe\,\textsc{xxv} triplet and the Fe\,\textsc{xxvi} doublet are resolved during egress, pre-ingress, and post-egress, whereas only one component of each is resolved during ingress.

These lines require a highly ionized plasma, with ionization parameters of the order of $\xi \sim L_X/(n r^2) \gtrsim 10^3$\,erg\,cm$^{-1}$ \citep{Kallman2004}. Radiative-transfer calculations for Cen\,X-3 using XSTAR and CLOUDY combined with a CAK wind density profile have shown that such highly ionized plasma can naturally arise in the photoionized stellar wind surrounding the NS \citep{Mochizuki2025_fex}. The presence of Fe\,\textsc{xxv} and \textsc{xxvi} emission across all phases constrains the emitting plasma to regions where the ionization parameter remains sufficiently high. From the definition of the ionization parameter,

\begin{equation}
r \lesssim \sqrt{\frac{L_X}{n \xi_\mathrm{min}}},
\end{equation}

where $L_X \sim 10^{37}$\,erg\,s$^{-1}$ is the typical X-ray luminosity of Cen\,X-3. Assuming a representative wind density of $n \sim 10^{11}$\,cm$^{-3}$ at the orbital separation, we obtain $r \lesssim$ a few solar radii. This scale is consistent with a compact photoionized region embedded in the stellar wind around the NS \citep{Mochizuki2025_fex}.

Ingress corresponds to a comparatively short interval of the orbit, and the
correspondingly lower photon statistics may limit the ability to resolve
individual components that are otherwise separable in energy during the
longer phases. 
We therefore conclude that the highly
ionized plasma is present throughout the orbit, including during ingress,
even though its substructure is not fully resolved. 

\subsection{Continuum variability and spectral state across the orbit}
\label{sec:disc_continuum}

In the two OOE intervals, the power-law photon indices are consistent within
($\Gamma \sim 1.5$), as are the normalizations
($\sim11.4 \times 10^{48}$\,ph\,s$^{-1}$\,keV$^{-1}$), indicating a relatively
stable accretion state outside eclipse. This is consistent with previous
NuSTAR spectral monitoring of Cen\,X-3, which finds little intrinsic
variability in the OOE photon index between observations at similar orbital
phases \citep{https://doi.org/10.48550/arxiv.2504.10670}.

The spectrum is hardest at egress, reaching $\Gamma \sim 1.1$, while the
photon index increases to $\Gamma \approx 1.4$ during ingress. The hard
spectrum at egress is consistent with the spectral behaviour typically
observed during eclipse transitions in previous X-ray studies of Cen\,X-3
\citep{2021MNRAS.501.5892S, Naik2011}, and may reflect the attenuation of the
softer part of the continuum by the residual stellar-limb column as the NS
emerges from eclipse.

At ingress, the photon index softens to $\Gamma \approx 1.4$, accompanied by
a factor ${\sim}1.5$ decrease in the power-law normalization relative to the
OOE phases. This apparent softening can be interpreted as a consequence
of the heavy absorbing column suppressing the direct hard continuum below
${\sim}7$\,keV: the observed emission increasingly reflects scattered and
reprocessed photons from the extended wind and disk, which contribute a softer
spectral component \citep{2003ApJ...582..959W, Naik2011}. 

\section{Conclusions}
\label{sec:conclusions}
We have presented \xrism\ spectroscopy of \cenx\ across four
orbital intervals (egress, post-egress, pre-ingress, and ingress),
providing the first high-resolution microcalorimeter study of both
eclipse transitions in an eclipsing HMXB. In particular, the
ingress and egress spectra offer a unique view of the changing
absorbing and reprocessing environment as the NS enters
and emerges from eclipse. Our main results are: 

\begin{enumerate}

\item {Orbital modulation of the local absorption.}
The local absorbing column varies by nearly two orders of magnitude over
the orbit, from
$N_{\rm H}=4.25\times10^{22}\,\mathrm{cm^{-2}}$ during pre-ingress to
$5.9\times10^{23}\,\mathrm{cm^{-2}}$ at ingress, corresponding to a
factor of $\sim3.4$ asymmetry between ingress and egress. A
spherically symmetric CAK wind model reproduces the egress and
out-of-eclipse column densities for a mass-loss rate of
$\dot{M}\simeq7.13\times10^{-7}\,M_\odot\,\mathrm{yr^{-1}}$, consistent
with previous estimates, but underpredicts the ingress absorption by
nearly an order of magnitude.

\item {Evidence for a compact overdense absorber.}
The excess absorbing column at ingress,
$\Delta N_{\rm H}\simeq3.3\times10^{23}\,\mathrm{cm^{-2}}$, is confined
to a narrow orbital-phase interval
($\Delta\phi\simeq0.014$), implying a compact structure with a
characteristic density of
$n_{\rm H}\sim2.8\times10^{12}\,\mathrm{cm^{-3}}$, approximately
300 times denser than the ambient stellar wind. This is consistent with
a localized accretion/photoionization wake or a disk-related overdensity,
rather than an extended wake spanning a large fraction of the orbit.

\item {Orbital evolution of the Fe\,K$\alpha$ velocity width.}
The Fe\,K$\alpha$ velocity width shows different best-fit values across
the orbit, with the broadest profile measured during egress,
$\sigma_{\rm Fe\,K\alpha}\approx1400\,\mathrm{km\,s^{-1}}$, and the
narrowest during pre-ingress,
$\sigma_{\rm Fe\,K\alpha}\approx510\,\mathrm{km\,s^{-1}}$. The ingress
and post-egress widths are statistically consistent, with
$\sigma_{\rm Fe\,K\alpha}\approx900\,\mathrm{km\,s^{-1}}$ in both
intervals. The observed differences suggest that the Fe\,K$\alpha$
fluorescence may arise from multiple reprocessing regions, with
broader emission associated with the dense environment around the NS
during the eclipse transitions and a narrower component becoming more
prominent before eclipse. Assuming Keplerian rotation around the NS,
the pre-ingress width corresponds to a reprocessing radius roughly a
factor of $\sim7$ larger than that implied by the broad egress
component.

\item {Orbital evolution of the Fe\,K$\alpha$ line flux.}
The Fe\,K$\alpha$ line flux varies across the orbit and shows no clear
correlation with the velocity width. In particular, ingress and
post-egress have statistically consistent widths but markedly
different line fluxes, indicating that the line intensity and
kinematic broadening are not directly coupled. The line flux is higher
during egress and post-egress, as the NS emerges from eclipse, and
lower during ingress and pre-ingress, as it approaches eclipse. The persistence of the
Fe\,K$\alpha$ emission suggests that an additional, more extended
fluorescent component remains visible throughout the orbit.

\item {Ionization structure of the Fe\,K complex.}
Highly ionized Fe\,\textsc{xxv} and Fe\,\textsc{xxvi} emission is detected
throughout the orbit. Several components of the
Fe\,\textsc{xxv} triplet and the Fe\,\textsc{xxvi} doublet are resolved
during egress, pre-ingress, and post-egress, whereas only one component
of each is resolved during ingress. This indicates that highly ionized
photoionized plasma surrounding the NS persists throughout the orbit,
including during the eclipse transitions. The required ionization state
is consistent with a compact photoionized region embedded in the stellar
wind, with a characteristic scale of a few solar radii for the adopted
luminosity and wind density.

\end{enumerate}

Together, these results demonstrate the unique capability of
XRISM/Resolve to disentangle the absorption, emission, and kinematic
properties of the Fe\,K complex, revealing a stratified and dynamically
complex circumstellar environment in Cen\,X-3 shaped by both the stellar
wind and a compact absorbing structure that
dominates immediately before eclipse ingress.

\begin{acknowledgements}
GSF, JPV, JJRR, and JMT acknowledge the financial support from the MICIU/AEI/10.13039/501100011033 with funding from the European Union (FEDER). Project (NewAthena24-UA), reference PID2024-155779OB-C33.
\end{acknowledgements}

\bibliographystyle{aa}
\bibliography{references}

\appendix
\onecolumn

\section{Parameter degeneracies}
\label{app:additional_analysis}

\begin{figure*}[h!]

      \centering

    \includegraphics[width=0.91\textwidth]{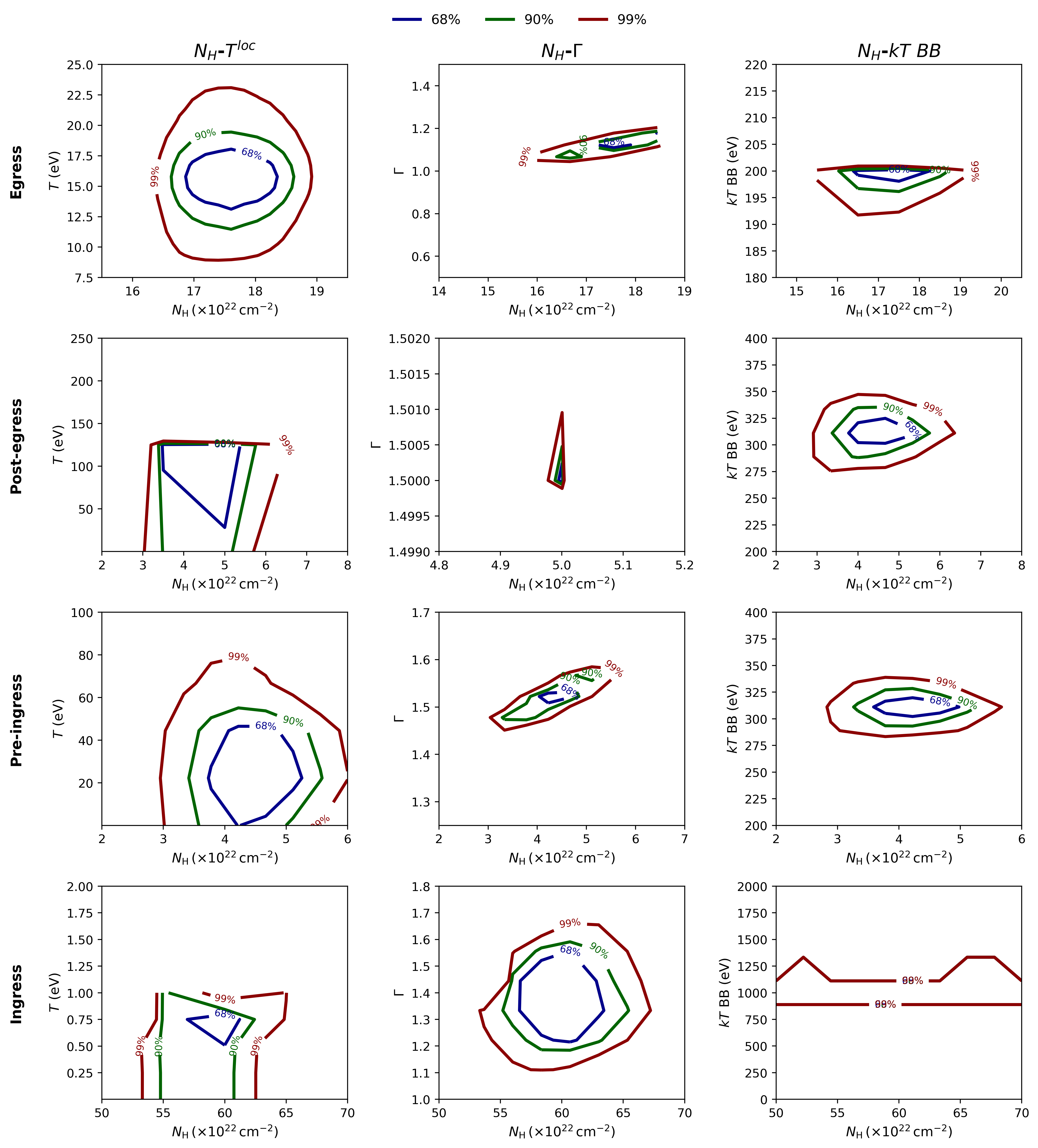}

    \caption{Joint confidence contours for
    $N_\mathrm{H}^\mathrm{local}$ versus $t^\mathrm{local}$ (left),
    the photon index $\Gamma$ (middle), and the disk blackbody
    temperature $kT$ (right). From top to bottom, the rows
    correspond to the egress, post-egress, pre-ingress, and ingress
    orbital intervals.}
    \label{fig:contour_plots}
\end{figure*}

To assess the impact of parameter covariance on the spectral
parameters discussed in the main text, we examined joint confidence
contours for the most relevant pairs of parameters
(Fig.~\ref{fig:contour_plots}): the local absorbing column density
$N_\mathrm{H}^\mathrm{local}$ versus the local absorber temperature
$t^\mathrm{local}$, the photon index $\Gamma$, and the disk
blackbody temperature $kT$.

The $N_\mathrm{H}^\mathrm{local}$--$t^\mathrm{local}$ contours show
that $t^\mathrm{local}$ becomes less constrained at the lowest
temperatures, while $N_\mathrm{H}^\mathrm{local}$ remains well
constrained. However, the contours do not extend towards
arbitrarily large $N_\mathrm{H}^\mathrm{local}$ as $t^\mathrm{local}$
approaches zero.

The $\Gamma$--$N_\mathrm{H}^\mathrm{local}$ contours indicate only weak
covariance for most orbital intervals, although some degree of
degeneracy is present during pre-ingress.

In contrast, the local absorbing column and the disk
blackbody temperature are more strongly correlated.
This degeneracy, together with the limited sensitivity of Resolve
below $\sim2$\,keV, means that the apparent orbital variations of the
disk blackbody component should be interpreted with caution.

\end{document}